\documentclass[12pt,aps,preprint,nofootinbib,superscriptaddress,nobalancelastpage]{revtex4}

\usepackage{mathrsfs}
\usepackage{amsmath,amsthm,amssymb,slashed}
\usepackage{color}
\usepackage{graphicx}
\usepackage{verbatim}
\usepackage{hyperref}
\usepackage{multirow}
\hypersetup{colorlinks,bookmarksopen,bookmarksnumbered,citecolor=blue,
linkcolor=blue,pdfstartview=FitH,urlcolor=blue}

\newcommand{\be}{\begin{equation}}
\newcommand{\ee}{\end{equation}}
\newcommand{\bea}{\begin{eqnarray}}
\newcommand{\eea}{\end{eqnarray}}

\begin{document}
\title{Completely general bounds on Non-Unitary leptonic mixing}

\author{Josu Hernandez-Garcia}
\email{josu.hernandez@uam.es}
\affiliation{Departamento de F\'isica Te\'orica, Universidad Aut\'onoma de Madrid, Cantoblanco E-28049 Madrid, Spain}
\affiliation{Instituto de F\'isica Te\'orica UAM/CSIC, Calle Nicol\'as Cabrera 13-15, Cantoblanco E-28049 Madrid, Spain}

\begin{abstract}
We derive constraints on the mixing of heavy right-handed neutrinos with the SM fields in the most general Seesaw scenario where the heavy neutrinos are integrated out. Among the electroweak and flavour observables included in the global fit, $\mu\rightarrow e\gamma$ sets the present strongest bound on the additional neutrino mixing, while in the future it will be dominated by $\mu-e$ conversion in nuclei. Increasing its sensitivity in future experiments could probe Non-Unitarity in Lepton Flavour Violating processes. Nevertheless, in order to determine completely model-independent constraints, we provide a second set of bounds derived through a global fit that does not include LFV observables. These indirect constraints on the off-diagonal elements come from the diagonal bounds through the Schwarz inequality.
\end{abstract}

\maketitle
\section{Introduction}

It is well known that neutrino masses are one of the most promising open windows to physics beyond the Standard Model (SM). Simply by adding extra heavy right-handed neutrinos to the SM particle content, neutrino masses arise in a simple and natural way in the so called Type-I Seesaw~\cite{Minkowski:1977sc,Mohapatra:1979ia,Yanagida:1979as,GellMann:1980vs}. The aim of our work \cite{Fernandez-Martinez:2016lgt} is constraining the additional neutrino mixing by using a set of electroweak (EW) and flavour observables. For some other recent works on this topic see references~\cite{Antusch:2014woa,Antusch:2015mia,Fernandez-Martinez:2015hxa}.\\
Once the new heavy states that we have added are integrated out, the SM-Seesaw theory that remains, and which is valid till a given energy scale $\Lambda$, can be considered as a low energy effective theory where the new phenomenology is encoded in a set of effective operators. The first one is the dim-5 Weinberg operator \cite{Weinberg:1979sa} that after EW symmetry breaking (EWSB) generates the masses of the light neutrinos $\hat{m}$:
\begin{equation}
\dfrac{c_{\alpha \beta}^\text{dim-5}}{\Lambda} \left(\overline{L^c}_\alpha\tilde{\phi}^*\right)\left(\tilde{\phi}^\dagger L_\beta\right)\xrightarrow{\text{EWSB}} \hat{m}=m_D^t M_N^{-1}m_D
\label{eq:Weinberg_op}
\end{equation}
where $\phi$ denotes the SM Higgs field and $M_N$ is the Majorana mass allowed for the right-handed neutrinos by the SM Gauge symmetry. Notice that since this operator violates Lepton number symmetry (L) by two units, light neutrino masses violate this accidental symmetry of the SM.\\
At dim-6 the only~\cite{Broncano:2002rw} operator that appears at three level is:
\begin{equation}
\frac{c_{\alpha \beta}^\text{dim-6}}{\Lambda^2} \left(\overline{L}_\alpha\tilde{\phi}\right)i\gamma^\mu\partial_\mu\left(\tilde{\phi}^\dagger L_\beta\right)\xrightarrow{\text{EWSB}} \eta=\frac{1}{2}m_D^\dagger M_N^{-2}m_D
\label{eq:dim6_op}
\end{equation}
which conserves $L$, and which after EWSB generates non-canonical kinetic terms among the active neutrinos of the SM. Then, after diagonalizing and normalizing the operator to bring the kinetic terms to its canonical form, it induces Non-Unitarity in the mixing matrix that appears in lepton charged current interactions. As a result, the Pontecorvo-Maki-Nakagawa-Sakata mixing matrix $U_\text{PMNS}$ is not going to be unitary and to stress that we call it $N$. The deviations of $N$ from unitarity are given by the $\eta$ mixing matrix in the following way:
\begin{equation}
N=\left(I-\eta\right)U_\text{PMNS}
\label{eq:N_matrix}
\end{equation}
where $\eta$ is related with the active-heavy neutrino mixing $\Theta$ as follows:
\begin{equation}
\eta=\frac{1}{2}\Theta\Theta^\dagger\quad\text{where}\quad\Theta=m_DM_N^{-1} \,.
\end{equation}
Since $\eta$ is Hermitian, Eq. \ref{eq:N_matrix} represents the most general parametrization~\cite{FernandezMartinez:2007ms} for $N$.\\

If the smallness of $\hat{m}$ (light neutrino masses) comes only from the suppression of the new heavy scale $M_N$, the mixing $\eta$ is going to be much more suppressed since it has two powers of the same heavy scale (compare Eq.~\ref{eq:Weinberg_op} and Eq.~\ref{eq:dim6_op}). Thus experimental verification turns extremely challenging. 
Alternatively, the smallness of light neutrino masses may naturally stem from an underlying approximate symmetry~\cite{Mohapatra:1986bd,Bernabeu:1987gr,Branco:1988ex,Buchmuller:1990du,Pilaftsis:1991ug,Dev:2012sg} of the theory instead of a huge hierarchy of masses. Since dim-5 operator violates $L$ if we impose an approximate Lepton number symmetry to the model, the operator becomes suppressed and as a result light neutrino masses become small. What is more, since dim-6 operator conserves $L$ and remains unsuppressed, the mixing can be arbitrarily large. This is known in the literature as Inverse\cite{Mohapatra:1986bd,Bernabeu:1987gr}/Linear~\cite{Malinsky:2005bi} Seesaw models.
In particular, the only~\cite{Kersten:2007vk,Abada:2007ux} Dirac and Majorana mass matrices that leads to an underlying $L$ symmetry are:
\begin{equation}
m_D = \frac{v_\text{EW}}{\sqrt{2}} \left(
\begin{array}{ccc}
Y_{Ne} & Y_{N\mu} & Y_{N\tau} \\ \epsilon_1 Y'_{Ne} & \epsilon_1 Y'_{N\mu} & \epsilon_1 Y'_{N\tau} \\ \epsilon_2 Y''_{Ne} & \epsilon_2 Y''_{N\mu} & \epsilon_2 Y''_{N\tau}
\end{array}
\right)
\qquad
\textrm{and}
\qquad
M_N = \left(
\begin{array}{ccc}
\mu_1 & \Lambda & \mu_3 \\ \Lambda & \mu_2 & \mu_4 \\ \mu_3 & \mu_4 & \Lambda'
\end{array}
\right), \label{eq:texture}
\end{equation}
with $\epsilon_i$ and $\mu_j$ small $\slashed{L}$ parameters. By setting all $\epsilon_i=0$ and $\mu_j=0$, $L$ symmetry is indeed recovered with the following $L$ assignments $L_e = L_\mu = L_\tau = L_1 = -L_2 = 1$ and $L_3 = 0$. As a result $\hat{m}=0$ (3 massless neutrinos in the $L$-conserving limit) but an arbitrarily large mixing $\eta$ are obtained. Upon switching on the $L$-violating parameters in Eq.~(\ref{eq:texture}), we end up with a small masses for the light neutrinos while the mixings are still arbitrarily large.
In this work we derive the bounds on a \textbf{completely} general Seesaw (G-SS) model where the SM is extended by an \textbf{arbitrary number} of right-handed neutrinos, all of them are \textbf{heavier} than $\Lambda_\text{EW}$. The mixing matrix $N$ is parametrized by the $3\times 3$ $\eta$ Hermitian matrix via Eq.~(\ref{eq:N_matrix}). Thus the Non-Unitarity of the PMNS matrix is given by 6 free elements. However, by using the Schwarz inequality:
\begin{equation}
\eta_{\alpha\beta}\leq\sqrt{\eta_{\alpha\alpha}\eta_{\beta\beta}}
\label{eq:Schwarz}
\end{equation}   
we can already set indirect constraints on the off-diagonal entries of the mixing matrix.  

\section{List of observables}

\begin{figure}
\centering
\includegraphics[width=0.75\textwidth]{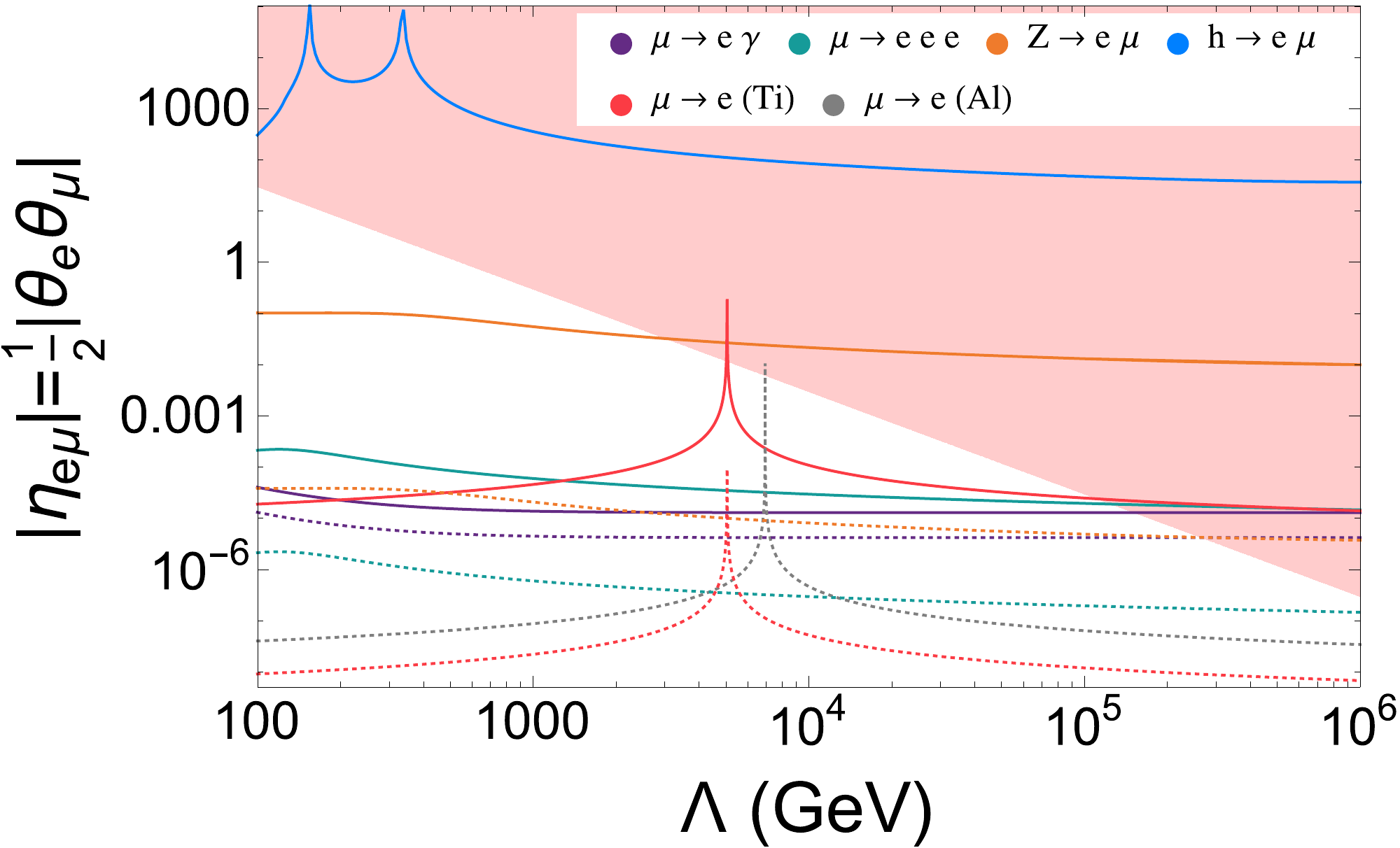}
\caption{$90$\% CL constraints on $\eta_{e\mu}$ from LFV 
observables. Solid lines represent current experimental bounds while dotted lines represent future 
sensitivities. The red-shadowed region represents the non-perturbative region with $|Y_N|^2>6\pi$.}
\label{fig:bounds}
\end{figure}

In this section I will briefly introduce the set of 28 EW and flavour observables we have used to constrain the additional neutrino mixing. The full expressions of the observables in terms of the Non-Unitarity parameters, $\alpha$, $G_\mu$ ($G_F$ measured in the $\mu$ decay) and $M_Z$ are given in the original paper \cite{Fernandez-Martinez:2016lgt}. The set of observables we have included in the global fit is:
\begin{itemize}
\item{The $W$ boson mass $M_W$}
\item{The effective weak mixing angle $\theta_\text{W}$: $s_\text{W eff}^{2 \text{ lep}}$ and $s_\text{W eff}^{2 \text{ had}}$}
\item{Four ratios of $Z$ fermionic decays: $R_l$, $R_c$, $R_b$ and $\sigma^0_\text{had}$}
\item{The invisible width of the $Z$ $\Gamma_\text{inv}$}
\item{Ratios of weak decays constraining EW universality: $R^\pi_{\mu e}$, $R^\pi_{\tau \mu}$, $R^W_{\mu e}$, $R^W_{\tau \mu}$, $R^K_{\mu e}$, $R^K_{\tau \mu}$, $R^l_{\mu e}$ and $R^l_{\tau \mu}$}
\item{9 weak decays constraining the CKM unitarity}
\item{3 radiative LFV decays: $\mu\rightarrow e \gamma$, $\tau\rightarrow \mu \gamma$ and $\tau\rightarrow e \gamma$}
\end{itemize}
We have studied in detail the Lepton Flavour Violating (LFV) decays which constraint the off-diagonal elements of the $\eta$ matrix. Since these processes become unsuppressed by the loss of the GIM cancellation we consider them worth investigating. A comparison summarizing the present relative importance of these observables constraining the off-diagonal elements of $\eta$ (solid lines) is presented in Fig.~\ref{fig:bounds}. As can be seen, the radiative decay $\mu \to e \gamma$ presently dominates. However, regarding future expectations (dotted lines), the constraints on $|\eta_{e \mu}|$ will be dominated by $\mu \to eee$ or $\mu-e$ conversion in nuclei rather than by $\mu \to e \gamma$. On the other hand, the present and future sensitivity to $|\eta_{e \tau}|$ and $|\eta_{\mu \tau}|$ is completely dominated by the radiative decays $\l_\alpha \to l_\beta \gamma$. In particular, the constraints on $|\eta_{\alpha\beta}|$ from the LFV decays of the $Z$ and Higgs bosons, $Z \to \l_\alpha\l_\beta$ and $h\to \l_\alpha\l_\beta$, are at least one or three orders of magnitude weaker than the bounds from radiative decays respectively.
For these reasons, the three radiative decays will thus be added to the global. However, since these processes depend on the heavy Majorana scale that is running in the loops, they could be considered as model-dependent processes. In such a way, we have performed two different global fits: the first one with the set of 28 observables which include the LFV decays, and the second one with just the 25 Lepton Flavour Conserving (LFC) observables that does not include the three rare decays.

\section{Results of the Global Fit}

\begin{figure}[htb!]
\centering
\includegraphics[width=0.45\textwidth]{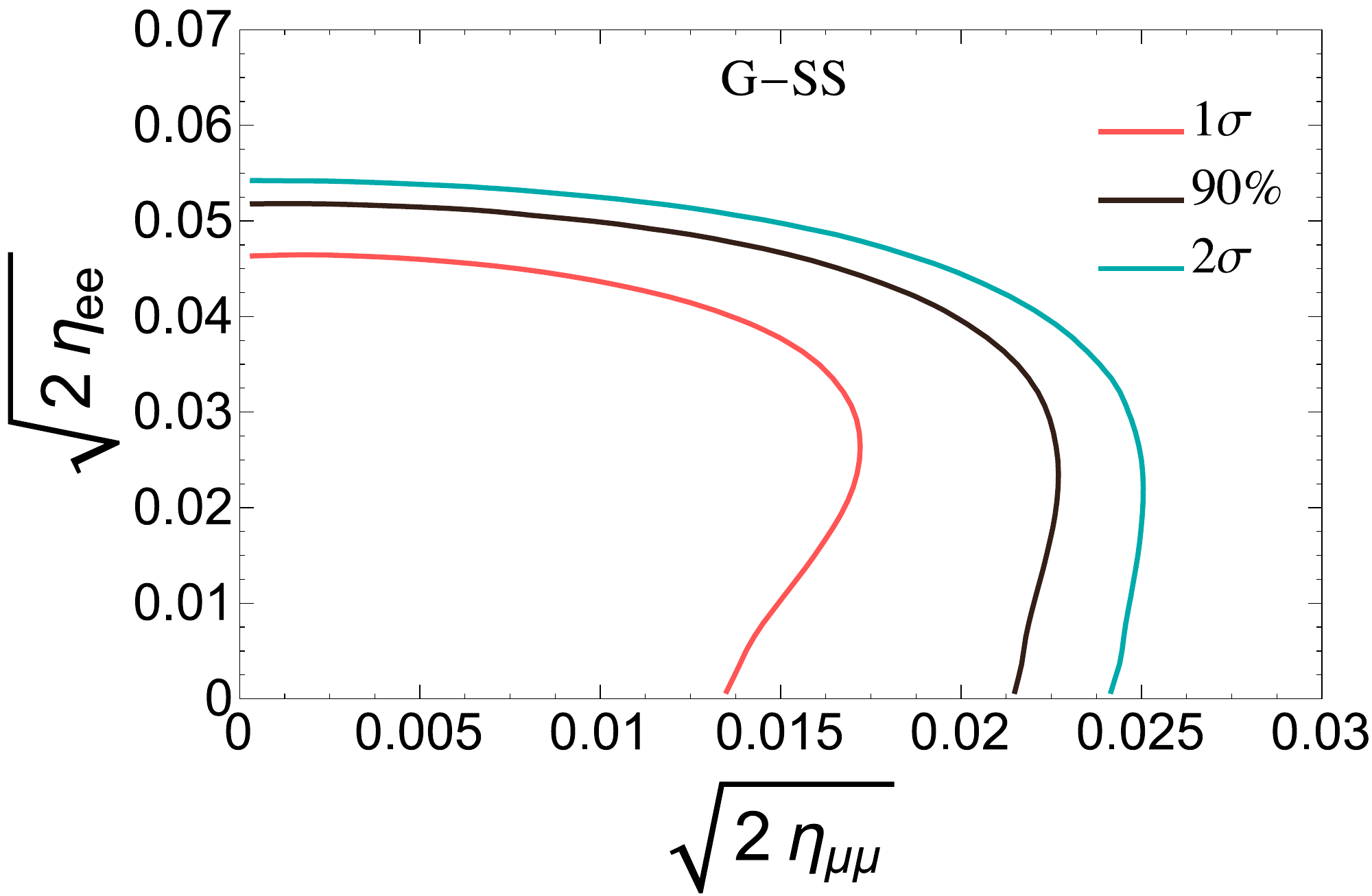}
\includegraphics[width=0.45\textwidth]{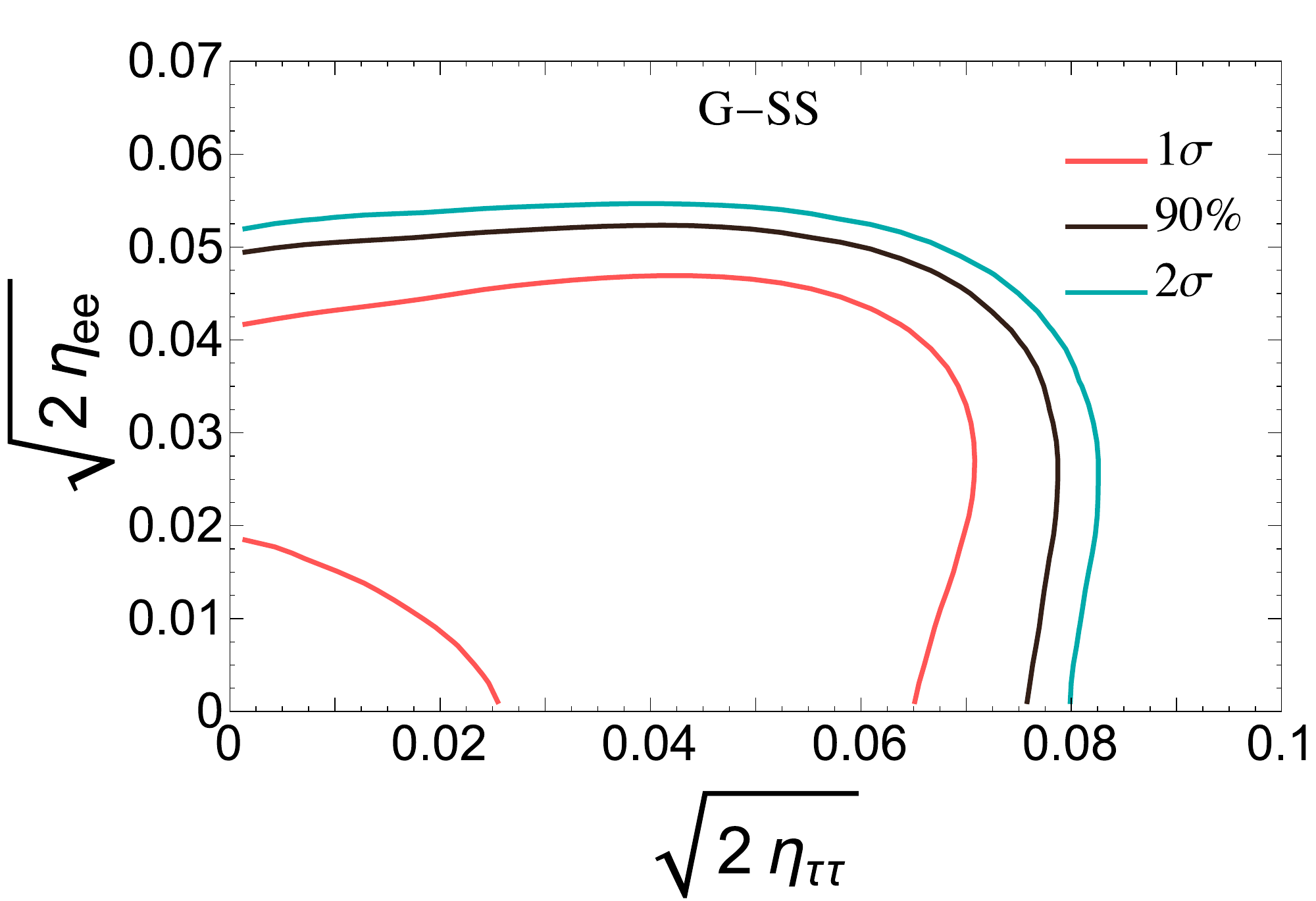}
\caption{Frequentist confidence intervals at $1 \sigma$, $90\%$ and $2 \sigma$ on the parameter space of the G-SS.}
\label{fig:contours}
\end{figure}

We have performed two separate Markov chain Monte Carlo (MCMC) simulations with the two sets of 28 and 25 observables that scan over the free parameters of the G-SS and as a result we obtain these frequentists contours at 1$\sigma$ (in red), 90$\%$ (in black) and 2$\sigma$ (in blue) showed in  Figure~\ref{fig:contours}.
Table~\ref{tab:bounds} summarizes the values of the Non-Unitarity parameters at $1\sigma$ and $2\sigma$ at the best fit point. These bounds are expressed in the quantity 
$\sqrt{2 |\eta_{\alpha \beta}|} = \sum_i \sqrt{ \Theta_{\alpha i} \Theta^*_{\beta i}}$. 
Thus, the diagonal elements $\sqrt{2 \eta_{\alpha \alpha}}$ correspond to the sum (in quadrature) of all mixings $\Theta_{\alpha i}$ of the individual extra heavy neutrinos $N_i$ to a given SM flavour $\alpha$ and represent an upper bound on each individual mixing. Concerning the diagonal bounds, we obtain a non-zero value with a significance close to $2\sigma$ for $\sqrt{2\eta_{ee}}$ and $\sqrt{2\eta_{\tau\tau}}$ while an upper bound for $\sqrt{2\eta_{\mu\mu}}$. Regarding the off-diagonal entries, the indirect bounds from LFC processes can be compared with the direct constraints from LFV observables. Interestingly, the constraint from $\mu \to e \gamma$ strongly dominates over all others leading to a bound one order of magnitude better $\sim 0.005$ in the $e-\mu$ entry, while the $e-\tau$ and $\mu-\tau$ values are rather dominated by the indirect constraints from the Schwarz inequality (comparison between the LFC and LFV rows).
\begin{table}[htb!]
\centering
\begin{tabular}{|c|c|c|c|c|c|c|c|}
\cline{3-8}
\multicolumn{2}{c|}{}&\multirow{2}{*}{$\sqrt{2\eta_{ee}}$}&\multirow{2}{*}{$\sqrt{2\eta_{\mu\mu}}$}&\multirow{2}{*}{$\sqrt{2\eta_{\tau\tau}}$}&\multirow{2}{*}{$\sqrt{2\eta_{e\mu}}$}&\multirow{2}{*}{$\sqrt{2\eta_{e \tau }}$}&\multirow{2}{*}{$\sqrt{2\eta_{\mu\tau}}$} \\
\multicolumn{2}{c|}{} &  &  &  &  &  &  \\
\hline
 \multirow{2}{*}{LFC} & $1\sigma$ & $\mathbf{0.031^{+0.010}_{-0.020}}$ & $\mathbf{< 0.011}$ &  $\mathbf{0.044^{+0.019}_{-0.027}}$ & $< 0.018$ & $\mathbf{< 0.045}$ & $\mathbf{< 0.024}$\\
 \cline{2-8}
 & $2\sigma$ & $\mathbf{< 0.050}$ & $\mathbf{< 0.021}$ & $\mathbf{< 0.075}$ & $< 0.026$ & $\mathbf{< 0.052}$ & $\mathbf{< 0.035}$\\
\hline
 \multirow{2}{*}{LFV} & $1\sigma$ & $-$ & $-$ & $-$ & $\mathbf{< 4.1 \cdot 10^{-3}}$ & $< 0.107$ & $< 0.115$\\
 \cline{2-8}
 & $2\sigma$ & $-$ & $-$ & $-$ & $\mathbf{< 4.9 \cdot 10^{-3}}$ & $< 0.127$ & $< 0.137$\\
 \hline
\end{tabular}
\caption{Comparison of all 1 and $2\sigma$ constraints on the heavy-active neutrino mixing. For the off-diagonal 
entries the indirect bounds from the LFC observables via the Schwarz inequality Eq.~(\ref{eq:Schwarz}) are compared with the direct LFV bounds and the dominant bound is highlighted in bold face.}
\label{tab:bounds}
\end{table}

\section{Conclusions}

We have used a set of EW and flavour observables to constrain the additional neutrino mixing of the most general Seesaw model. We have obtained a non-zero value for the Non-Unitarity parameters $\eta_{ee}$ and $\eta_{\tau \tau}$ with a significance close to $2\sigma$ and and upper bound for the $\eta_{\mu\mu}$ element. Concerning the off-diagonal elements when the LFV decays are included in the Global Fit $\eta_{\mu e}$ is dominated by $\mu\rightarrow e\gamma$ while $\eta_{\mu\tau}$ and $\eta_{e\tau}$ already get an stronger indirect constrain from the diagonal bounds via the Schwarz inequality. As a final remark, the updated bounds presented in the original paper\cite{Fernandez-Martinez:2016lgt} and collected in these pages apply to any extension of the SM with right-handed neutrinos heavier than the EW scale.

\begin{acknowledgments}
I acknowledge financial support by the European Union through the ITN ELUSIVES H2020-MSCA-ITN-2015//674896 and the RISE INVISIBLESPLUS H2020-MSCA-RISE-2015//690575. I also acknowledge support from the EU through the FP7 Marie Curie Actions CIG NeuProbes PCIG11-GA-2012-321582 and the Spanish MINECO through the ``Ram\'on y Cajal'' programme (RYC2011-07710), the HPC-Hydra cluster at IFT, the project FPA2012-31880 and through the Centro de excelencia Severo Ochoa Program under grant SEV-2012-0249.
\end{acknowledgments}

\end{document}